\documentclass{elsart}

\usepackage{amssymb}
\usepackage{graphicx,psfig,epsfig}

\def\la{\langle}
\def\ra{\rangle}
\def\beq{\begin{equation}}
\def\eeq{\end{equation}}
\def\be{\begin{eqnarray}}
\def\ee{\end{eqnarray}}
\def\hs{\hat{s}}
\def\htm{\hat{t}}
\def\hu{\hat{u}}

\def\k2av{\la k_T^2\ra}
\newcommand{\f}[2]{\frac{#1}{#2}}
\newcommand{\dd}{ {\textrm d}}

\begin{document}

\begin{frontmatter}



\title{Nuclear effects in dAu collisions from recent RHIC 
data\thanksref{1}}

\thanks[1]{Supported in part by U.S. DE-FG02-86ER40251, U.S. NSF INT-0435701, 
and Hungarian OTKA grants T043455, T047050, and NK062044.}


\author[a,b]{P. L\'evai}, \author[a,c]{G.G. Barnaf\"oldi}, 
\author[b]{G. Fai},  and \author[d]{G. Papp}

\address[a]{KFKI RMKI, 
            P.O. Box 49, Budapest H-1525, Hungary}
\address[b]{CNR, Department of Physics, Kent State University, Kent OH-44242, USA}
\address[c]{Department for Physics of Complex Systems, E{\"o}tv{\"o}s University, \\
	    P{\'a}zm{\'a}ny P. 1/A, Budapest 1117, Hungary}  
\address[d]{Department for Theoretical Physics, E{\"o}tv{\"o}s University, \\ 
	    P{\'a}zm{\'a}ny P. 1/A, Budapest 1117, Hungary}  

\begin{abstract}
Neutral pion ($\pi^0$) production is calculated in a leading
order (LO) perturbative QCD-based model in $pp$ and $dAu$ collisions
at $\sqrt{s}=200 $ AGeV at midrapidity. The model includes
the transverse component of the initial parton distribution.
We compare our results for $pp$ collision
to experimental data at RHIC energy. We repeat our calculation
for the $dAu$ collision and investigate
the interplay between shadowing and multiple scattering.
In central $dAu$ collisions the influence of possible jet energy loss on cold nuclear
matter is discussed and numerical results are displayed.

\end{abstract}

\end{frontmatter}

\vspace*{-0.3cm}
\section{Introduction}
\label{sec_int}
\vspace*{-0.4cm}

Experimental data on $\pi^0$ production
in $Au+Au$ collisions at \mbox{$\sqrt{s}=$ 200 AGeV}
display a very strong suppression pattern
in the transverse momentum region 
\mbox{2 GeV $< p_T <$ 20 GeV~\cite{PHENpi0200}},
which indicates a clear evidence
of the presence of induced jet energy loss in hot dense 
matter~\cite{gptw,GLV,BDMS00,Lev02}.
Quantitative calculations of the energy loss in $Au+Au$ collisions
require the precise description of the available data from $d+Au$ collisions. These
data incorporate initial state effects (nuclear multiscattering 
and nuclear shadowing) and possible final state effects (e.g. energy loss
in cold nuclear matter), which should be understood. 

\section{The parton model for $pp$ collisions}
\label{sec_mod}
\vspace*{-0.3cm}
In this paper we accomplish the analysis of $pp$ and $dAu$ data in a leading-order (LO)
perturbative QCD based model~\cite{Wang01,YZ02}. We plan to repeat our analysis in
next-to-leading (NLO) order~\cite{BGG04}, when final data will be available for
pion production in $dAu$ collisions.
The invariant cross section for pion production in
$pp$ collision can be described in a LO pQCD-improved parton model
on the basis of the factorization
theorem as a convolution~\cite{Wang01,YZ02}:
\begin{eqnarray}
\label{hadpp}
  E_{\pi^0}\f{\dd \sigma_{\pi^0}^{pp}}{\dd ^3p} &=&
        \sum_{abcd} \int\! \dd x_a \dd x_b \dd z_c 
        \ f_{a/p}(x_a,{\bf k}_{Ta},Q^2)
        \ f_{b/p}(x_b,{\bf k}_{Tb},Q^2) \cdot \ \nonumber \\
&&  \ \ \ \ \cdot \f{\dd \sigma}{\dd \htm}(ab \to cd)\,
   \frac{D_{\pi^0/c}(z_c, Q'^2)}{\pi z_c^2} \, \hs \, \delta(\hs+\htm+\hu)\ \ \  ,
\end{eqnarray}
where  
$\dd \sigma/ \dd\htm$ is the hard scattering cross section of the
partonic subprocess $ab \to cd$,  and the 
fragmentation function (FF), $D_{h/c}(z_c, Q'^2)$
gives the probability for parton $c$ to fragment into hadron $h$
with momentum fraction $z_c$ at scale $Q'$.
For fragmentation functions we use the 
KKP parameterization~\cite{KKP}.

The functions
$f_{a/p}(x,{\bf k}_{Ta},Q^2)$ and  
$f_{b/p}(x,{\bf k}_{Tb},Q^2)$  are the 
parton distribution functions (PDFs) for the
colliding partons $a$ and $b$ in the interacting protons
as functions of momentum fraction $x$, at scale $Q$. 
We use a product approximation, 
namely $f(x,{\bf k}_{T},Q^2) = f(x,Q^2) g({\bf k}_{T})$, where
the function $f(x,Q^2)$ represents the standard LO PDF 
(here it is the GRV set~\cite{GRVLO}).
In our phenomenological approach the transverse-momentum distribution
is described by a Gaussian,
\beq
\label{kTgauss}
g({\bf k}_T) \ = \f{1}{\pi \la k^2_T \ra}
        \  \exp \left(-\frac{k^2_T}{\la k^2_T \ra} \right)    \,\,\, .
\eeq
Here, $\langle k_T^2 \rangle$ is the 2-dimensional width of the $k_T$
distribution and it is related to the magnitude of the
average transverse momentum of a parton
as $\langle k_T^2 \rangle = 4 \langle k_T \rangle^2 /\pi$.

The validity of this approximation and the introduction of the
transverse component into the PDFs were under debate for a long time.
However, recent measurement of di-hadron correlations by 
PHENIX~\cite{dihadronPHENIX} and the theoretical investigation of these
data~\cite{dihadrontheo} have shown that a large value for the 
$k_T$-imbalance can be extracted from existing di-hadron data. The origin
of this $k_T$-imbalance is three-fold: intrinsic transverse
momenta of the partons, non-perturbative soft radiation and
higher order radiation terms. We consider these effects phenomenologically
and our $\langle k_T^2 \rangle$ value accumulates all of these effects.
As we will see, the size of this $\langle k_T^2 \rangle$ can be 
extracted consistently from one-particle spectra and two-particle 
correlations.

In our present study we consider fixed scales:
the factorization scale is connected to the momentum of the intermediate jet,
$Q=\kappa\cdot p_q$ (where $p_q=p_T/z_c$),
while the fragmentation scale is connected to the final hadron momentum,
$Q_F=\kappa \cdot p_T$.
The value of $\kappa$ can be varied in a wide range, however we will
use $\kappa \ = \ 2/3$, which is the best choice to
obtain agreement between data and theory at high-$p_T$, where non-perturbative
effects have a small contribution.

\begin{figure}[h]
\begin{minipage}{88mm}
\vspace*{-0.2cm}
\resizebox{88mm}{100mm}
{\includegraphics{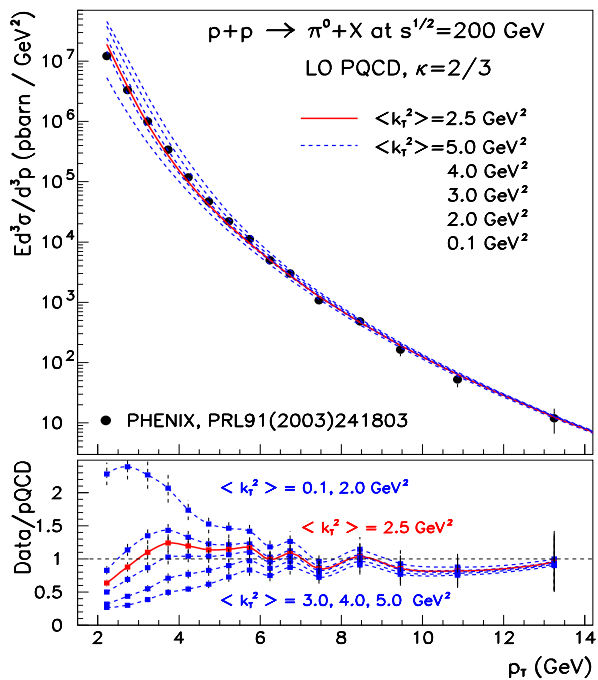}}
\vspace*{-0.8cm}
\caption{\footnotesize
(Color online) One-particle spectra for $\pi^0$ production 
in $pp$ collisions at $\sqrt{s} = 200$~GeV at different intrinsic $k_T$ values
compared to PHENIX data~\cite{PHENpi0200pp}. 
}
\label{fig1}
\end{minipage}
\hspace{\fill}
\begin{minipage}{49mm}
{Figure 1 displays our results at $\sqrt{s}=200$ GeV
for $pp\rightarrow \pi^0X$~\cite{PHENpi0200pp}
at different values of the $k_T-$imbalance.
Without transverse momentum component for partons
the theoretical results underestimate the data by a factor of 2
in the window  2 GeV $< p_T <$ 5 GeV.
To minimize the difference
between RHIC data and the model, a value of
$\langle k_T^2 \rangle = 2.5$ GeV$^2$ can be selected.
This value indicates 
\mbox{$\langle p_{T,pair}^2 \rangle = 2 \langle k_T^2 \rangle
= 5$ GeV$^2$,} which is compatible with the value obtained
in Refs.~\cite{dihadronPHENIX,dihadrontheo} from
di-hadron spectra. }
\end{minipage}
\end{figure}

Our earlier studies have shown that the theoretical reproduction
of the one-hadron spectra in a pQCD frame strongly requires the
introduction of a large $k_T-$imbalance in both LO and NLO 
calculations~\cite{Levai03}. The analysis of di-hadron
correlations in  NLO frame would give more insight into
this problem.

\section{The parton model for $dAu$ collisions}

Considering the $dAu$ collision, the hard pion production cross section
can be written as an integral over impact parameter $b$, where
the geometry of the collision is  described in
the Glauber picture~\cite{Wang01,YZ02}:
\beq
\label{dAuX}
  E_{\pi}\f{\dd \sigma_{\pi}^{dAu}}{ \dd ^3p} =
  \int \dd ^2b \, \dd ^2r \,\, t_d(r) \,\, t_{Au}(|{\bf b} - {\bf r}|) \cdot
  E_{\pi} \,    \f{\dd \sigma_{\pi}^{pp}(\k2av_{pAu},\k2av_{pd})}
{\dd ^3p}
\,\,\, ,
\eeq
where the proton-proton cross section on the right hand side represents
the cross section from eq.~(\ref{hadpp}), but with the
broadened widths of the transverse-momentum distributions 
in eq.~(\ref{kTgauss}), as a consequence
of nuclear multiscattering (see eq. (\ref{ktbroadpA})).
In eq.~(\ref{dAuX}), $t_{Au}(b) = \int \dd z \, \rho_{Au}(b,z)$
is the nuclear thickness function (in terms of the density distribution
of the gold nucleus, $\rho_{Au}$),
normalized as $\int \dd ^2b \, t_{Au}(b) = A_{Au} = 197$.
For the deuteron, one could use a superposition of a $pAu$ and a $nAu$
collision, or a distribution for the nucleons inside the deuteron.
Here we apply a hard-sphere approximation for the deuteron with A=2 for
estimating the nuclear effects. Also, since $\pi^0$ production
is not sensitive to isospin, we continue to use the notation ``$pA$''
when talking about the interaction of any nucleon with a nucleus.

The initial state broadening of the incoming parton
distribution function is accounted for by an
increase in the width of the
gaussian parton transverse momentum distribution 
in eq. (\ref{kTgauss}) \cite{YZ02,Papp02}:
\beq
\label{ktbroadpA}
\k2av_{pA} = \k2av_{pp} + C \cdot h_{pA}(b) \ .
\eeq
Here, $\k2av_{pp}$ is the width of the transverse momentum distribution
of partons in $pp$ collisions,
$h_{pA}(b)$ describes the number of {\it effective}
nucleon-nucleon (NN) collisions at impact parameter $b$,
which impart an average transverse momentum squared $C$.
The effectivity function $h_{pA}(b)$ can be written in terms of the
number of collisions suffered by the incoming proton in the target
nucleus, $h_{pA}(b)= \nu_A(b)-1$. Here, $\nu_A(b) = \sigma_{NN} t_{A}(b)$,
with $\sigma_{NN}$ being
the inelastic nucleon-nucleon cross section.
For the factor $C$ and $\nu_A(b)$ we will use the findings of
Ref.~\cite{YZ02}, where the systematic analysis of $pA$ reactions
was performed in LO and the characteristics
of the Cronin effect were determined  at LO level.
Following Ref.~\cite{YZ02}, we
assume that only a limited number of semi-hard collisions
(with maximum  $\nu_A(b)_{max} = 4$)
contributes to the broadening, and we have found $C =$ 0.4 GeV$^2$. 
The maximum of the broadening is \mbox{$\sim $1  GeV$^2$} and
this value determines the maximum increase in the particle yield.

It is well-known that the PDFs are modified in the
nuclear environment and nuclear shadowing appears for partons 
with momentum fraction $x < 0.1$. This is taken into account
by various shadowing parameterizations~\cite{EKS,Shadxnw_uj,HKM,Fran02}.
In the present work, we display results obtained with the
EKS parameterization, which has an antishadowing
feature\cite{EKS}, and with
the updated HIJING parameterization\cite{Shadxnw_uj}, which
incorporates different quark and gluon shadowing and requires
the introduction of nuclear multiscattering.
As a third case, we consider the nuclear parton distribution
functions introduced by HKM~\cite{HKM}. Since beyond nuclear shadowing
the nuclear multiscattering may appear, we will apply the HKM parametrization
alone and in a combination with our multiscattering description.

\section{ Results on pion production in $dAu$ collisions}

Including the multiscattering and shadowing effects one can
calculate the invariant cross section for pion production
in $d+Au$ collision.
Moreover, introducing the nuclear modification factor
$R_{dAu}$, as
\be
\label{rdau}
R_{dAu} = \f{E_{\pi}\dd \sigma_{\pi}^{dAu}/\dd ^3p}
            {N_{bin} \cdot E_{\pi}\dd \sigma_{\pi}^{pp}/\dd ^3p}
= \f{E_{\pi}\dd \sigma_{\pi}^{dAu}({\tt \ with \ nuclear \ effects})/\dd ^3p}
{E_{\pi}\dd \sigma_{\pi}^{dAu}({\tt \ no \ nuclear \ effects})/\dd ^3p} 
\,\, , \, \, \  
\ee
nuclear effects can be investigated clearly and efficiently, using
a linear scale. The value of $N_{bin}$ can be determined using the Glauber
geometrical overlap integral as in eq. (\ref{dAuX}). Here we apply
the right-most equation in our theoretical calculations, 
which does not  require the determination of
$N_{bin}$ from the Glauber model.

\vspace*{-0.4cm}
\begin{figure}[h]
\begin{minipage}{139mm}
\vspace*{-0.2cm}
\resizebox{148mm}{120mm}
{\includegraphics{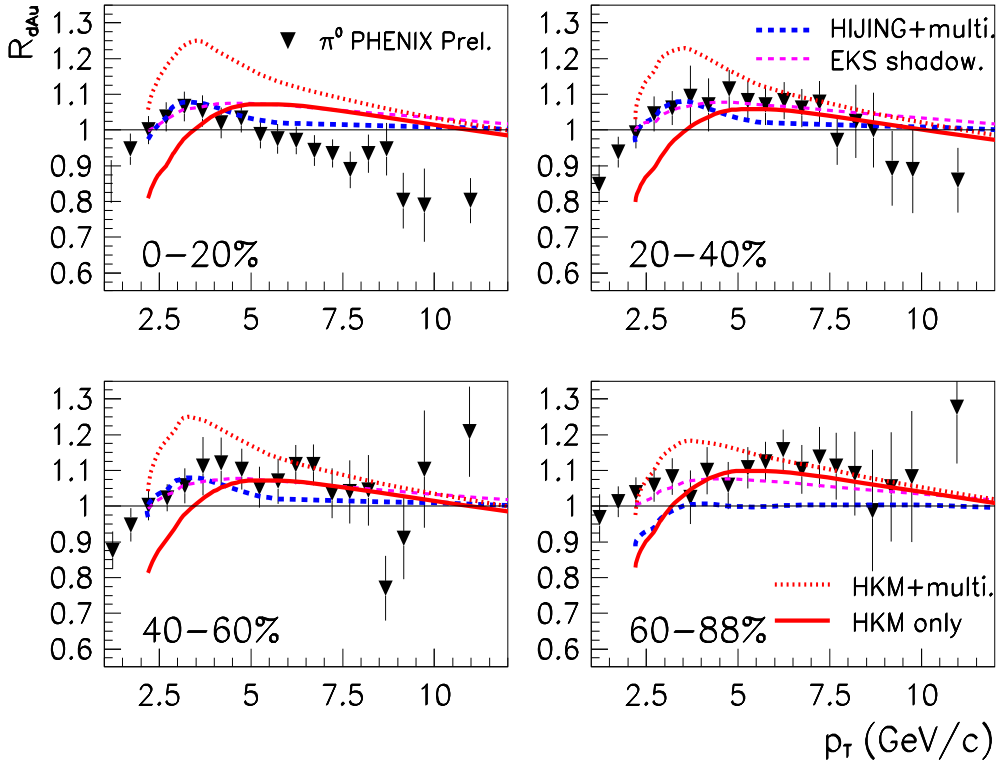}}
\vspace*{-0.8cm}
\caption{\footnotesize
(Color online) Nuclear modification factor for $\pi^0$ production
in $dAu$ collisions at $\sqrt{s_{NN}} = 200$~AGeV at different centralities.
Theoretical results (see text for details) are compared 
to preliminary PHENIX data~\cite{PHENdAu200}.
}
\label{fig2}
\end{minipage}
\hspace{\fill}
\end{figure}

Figure 2 summarizes our theoretical results in four centrality bins
for the nuclear modification factor $R_{dAu}$
in $dAu$ collisions at $\sqrt{s}=200$ AGeV,
displaying different shadowing models. 
Preliminary PHENIX data are from Ref.~\cite{PHENdAu200}.

At first glance, 
the experimental data have a striking message at all centrality bins:
the nuclear modification factor shows a $\pm 10 \%$ deviation from 
unity in wide momentum regions in all centrality bins, including the
most central and the most peripheral ones. Since these
$\pm 10 \%$ deviations overlap with the systematic error
of these experimental data, we can not exclude the scenario of a null effect, 
which means no multiscattering and no shadowing are present at RHIC energies.
Fortunately the nuclear modification factor has been measured at
higher rapidities and the influence of nuclear shadowing 
can be seen clearly~\cite{BGG0406},
which excludes the null effect scenario.
This means that there is a delicate interplay
between nuclear shadowing and multiscattering (or the antishadowing
feature of the shadowing functions).

Investigating the experimental data from this point of view, we may
recognize a slight b-dependence for both multiscattering and shadowing.
In the most peripheral collisions (60-88 \%) multiscattering 
overwhelms the shadowing - or shadowing function has a clear
antishadowing region to be involved in the pion production
in the region $3 < p_T < 7$ GeV/c. The enhancement can be reproduced by
the EKS (thin dashed line) and the 
HKM shadowing  (solid line), 
the scenario of 'HKM+multiscattering' (dotted line)
is satisfactory only because of the large error bars.
HIJING (thick dashed line) displays no nuclear modification
--- more precise data may help us to
choose between the different scenarios.
In the most central collisions (0-20 \%) 
in the intermediate transverse momentum window the $+ 10 \%$ enhancement
appears through the  superposition of a
stronger shadowing and a more effective multiscattering connected
to the geometry of the collisions. If final data will have a
smaller error bars, then we may be able to select between the different theoretical
models. 

However, data in the most central bins (0-20 \%) display an interesting phenomenon
at high-$p_T$: although all model calculations with nuclear multiscattering and 
shadowing lead to unity, the experimental data are decreasing with increasing
transverse momentum. The authors are not aware of any shadowing parametrization,
which could yield such a suppression around $p_T=6-10$ GeV at RHIC energies,
and any nuclear multiscattering should work in the opposite direction. This
suppression in central $dAu$ collisions may indicate the presence of a new
phenomena, namely a moderate jet energy loss in the cold matter produced
in $dAu$ collisions. Cold quenching of high energy quarks
is known from the HERMES experiment~\cite{HERMES}, and recently it was
investigated theoretically (see Refs.~\cite{EWangXN,Arleo03,Accardi06}).
The question of cold quenching is very complicated, and it is strongly 
related to the microscopical description of hadronization 
(see Ref.~\cite{Accardi06}).
Thus recent $dAu$ data with large error bars can not give a deep insight into
this phenomena beyond indicating its existence in the most central bins.
We estimated this effect with a simple rescaling
of the 'HKM+multiscattering' scenario. Figure 3 indicates that 
$dAu$ data in the most central bin 
can accommodate a 10-15 \% suppression.

\vspace*{-0.4cm}
\begin{figure}[h]
\begin{minipage}{139mm}
\vspace*{-0.2cm}
\resizebox{115mm}{88mm}
{\includegraphics{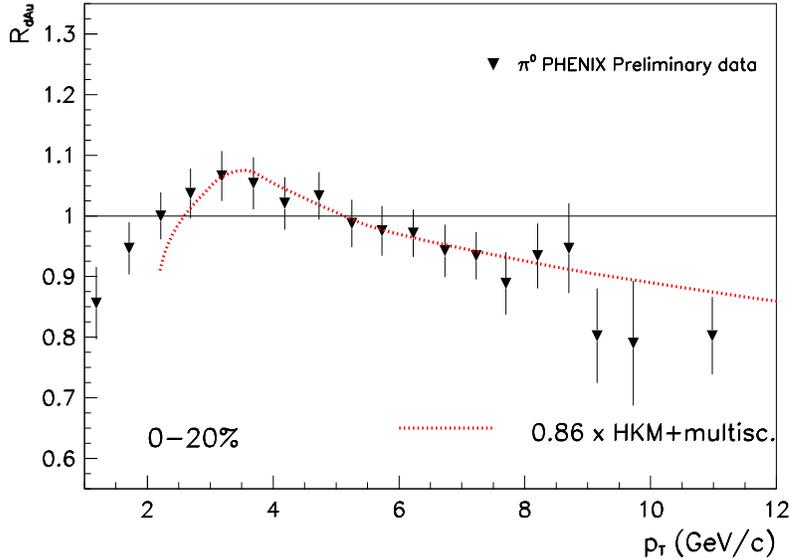}}
\caption{\footnotesize
(Color online) Nuclear modification factor for $\pi^0$ production
in central $dAu$ collisions at $\sqrt{s_{NN}} = 200$~AGeV.
Theoretical result (dotted line) is obtained by simple rescaling
of the 'HKM+multiscattering' scenario (see text for details).
Preliminary PHENIX data are from Ref.~\cite{PHENdAu200}.
}
\label{fig3}
\end{minipage}
\hspace{\fill}
\end{figure}

\section{Summary}

We have investigated the interplay between nuclear shadowing and
multiscattering at RHIC energies in $dAu$ collisions. Since the nuclear 
modification factor, $R_{dAu}$, has a value close to unity in large
transverse momentum regions, the two effects seem to balance each other
within $\pm 15 \%$, which makes the separation and the quantitative
analysis of these effects in the mid-rapidity difficult. 
Final data with smaller error bars may lessen
this complication. The characteristic suppression of the nuclear modification
factor at high-$p_T$ in the most central $dAu$ collisions may indicate the
presence of a moderate jet energy loss.
Quantitative analysis requires the introduction of proper theoretical
models for jet energy loss in cold nuclear matter.

\end{document}